\documentstyle[12pt]{article}

\catcode`\@=11
\long\def\@makefntext#1{
\protect\noindent \hbox to 3.2pt {\hskip-.9pt
$^{{\ninerm\@thefnmark}}$\hfil}#1\hfill}		

\def\@makefnmark{\hbox to 0pt{$^{\@thefnmark}$\hss}}  

\def\ps@myheadings{\let\@mkboth\@gobbletwo
\def\@oddhead{\hbox{}
\rightmark\hfil\ninerm\thepage}
\def\@oddfoot{}\def\@evenhead{\ninerm\thepage\hfil
\leftmark\hbox{}}\def\@evenfoot{}
\def\sectionmark##1{}\def\subsectionmark##1{}}

\renewcommand{\thefootnote}{\fnsymbol{footnote}}

\newcounter{sectionc}\newcounter{subsectionc}\newcounter{subsubsectionc}
\renewcommand{\section}[1] {\vspace*{0.6cm}\addtocounter{sectionc}{1}
\setcounter{subsectionc}{0}\setcounter{subsubsectionc}{0}\noindent
	{\normalsize\bf\thesectionc. #1}\par\vspace*{0.4cm}}
\renewcommand{\subsection}[1] {\vspace*{0.6cm}\addtocounter{subsectionc}{1}
	\setcounter{subsubsectionc}{0}\noindent
	{\normalsize\it\thesectionc.\thesubsectionc. #1}\par\vspace*{0.4cm}}
\renewcommand{\subsubsection}[1]
{\vspace*{0.6cm}\addtocounter{subsubsectionc}{1}
	\noindent {\normalsize\rm\thesectionc.\thesubsectionc.\thesubsubsectionc.
	#1}\par\vspace*{0.4cm}}

\newcounter{appendixc}
\newcounter{subappendixc}[appendixc]
\newcounter{subsubappendixc}[subappendixc]

\renewcommand{\appendix}[1] {\vspace*{0.6cm}
        \refstepcounter{appendixc}
        \setcounter{figure}{0}
        \setcounter{table}{0}
        \setcounter{equation}{0}
        \renewcommand{\thefigure}{\Alph{appendixc}.\arabic{figure}}
        \renewcommand{\thetable}{\Alph{appendixc}.\arabic{table}}
        \renewcommand{\theappendixc}{\Alph{appendixc}}
        \renewcommand{\theequation}{\Alph{appendixc}.\arabic{equation}}
        \noindent{\bf Appendix \theappendixc #1}\par\vspace*{0.4cm}}

\def\abstracts#1{{

\centering{\begin{minipage}{12.2truecm}\footnotesize\baselineskip=12pt\noindent
	\centerline{\footnotesize ABSTRACT}\vspace*{0.3cm}
	\parindent=0pt #1
	\end{minipage}}\par}}

\renewenvironment{thebibliography}[1]
	{\begin{list}{\arabic{enumi}.}
	{\usecounter{enumi}\setlength{\parsep}{0pt}
\setlength{\leftmargin 1.25cm}{\rightmargin 0pt}
	 \setlength{\itemsep}{0pt} \settowidth
	{\labelwidth}{#1.}\sloppy}}{\end{list}}

\newcounter{itemlistc}
\newcounter{romanlistc}
\newcounter{alphlistc}
\newcounter{arabiclistc}

\newcommand{\fcaption}[1]{
        \refstepcounter{figure}
        \setbox\@tempboxa = \hbox{\footnotesize Fig.~\thefigure. #1}
        \ifdim \wd\@tempboxa > 6in
           {\begin{center}
        \parbox{6in}{\footnotesize\baselineskip=12pt Fig.~\thefigure. #1}
            \end{center}}
        \else
             {\begin{center}
             {\footnotesize Fig.~\thefigure. #1}
              \end{center}}
        \fi}

\newcommand{\tcaption}[1]{
        \refstepcounter{table}
        \setbox\@tempboxa = \hbox{\footnotesize Table~\thetable. #1}
        \ifdim \wd\@tempboxa > 6in
           {\begin{center}
        \parbox{6in}{\footnotesize\baselineskip=12pt Table~\thetable. #1}
            \end{center}}
        \else
             {\begin{center}
             {\footnotesize Table~\thetable. #1}
              \end{center}}
        \fi}

\def\@citex[#1]#2{\if@filesw\immediate\write\@auxout
	{\string\citation{#2}}\fi
\def\@citea{}\@cite{\@for\@citeb:=#2\do
	{\@citea\def\@citea{,}\@ifundefined
	{b@\@citeb}{{\bf ?}\@warning
	{Citation `\@citeb' on page \thepage \space undefined}}
	{\csname b@\@citeb\endcsname}}}{#1}}

\newif\if@cghi
\def\cite{\@cghitrue\@ifnextchar [{\@tempswatrue
	\@citex}{\@tempswafalse\@citex[]}}
\def\citelow{\@cghifalse\@ifnextchar [{\@tempswatrue
	\@citex}{\@tempswafalse\@citex[]}}
\def\@cite#1#2{{$\null^{#1}$\if@tempswa\typeout
	{IJCGA warning: optional citation argument
	ignored: `#2'} \fi}}

 1
 1
 1

\font\ninerm=cmr9

\begin{document}
\centerline{\normalsize \hspace{6in} LSUHE No. 184-1995}
\centerline{\normalsize\bf DUAL ABRIKOSOV VORTICES IN CONFINING THEORIES}
\baselineskip=22pt

\centerline{\footnotesize Richard W. Haymaker}
\baselineskip=13pt
\centerline{\footnotesize\it Department of Physics and Astronomy,
Louisiana State University}
\baselineskip=12pt
\centerline{\footnotesize\it Baton Rouge, Louisiana, 70803, USA}
\centerline{\footnotesize E-mail: haymaker@rouge.phys.lsu.edu}
\vspace*{0.3cm}

\vspace*{0.9cm}
\abstracts{The spacial distribution of fields and currents
in confining theories can give direct evidence of dual
superconductivity.  We
would like to discuss the techniques for finding these properties
and calculating the superconductivity parameters in lattice
simulations.  We have seen dual Abrikosov vortices directly in pure
U(1) and SU(2) and others have also seen them in SU(3).  In the non-Abelian
cases the system appears to be near the borderline between type I and II.
We also discuss the response of the supercurrents to external fields.}

\normalsize\baselineskip=15pt
\setcounter{footnote}{0}
\renewcommand{\thefootnote}{\alph{footnote}}
\section{Introduction}

Lattice gauge theory offers the prospect of
exploring dual superconductivity\cite{back1}
 in depth as a confining mechanism.
The existence of a dual Abrikosov vortex
between a static quark-antiquark pair leads to a linearly rising potential
energy between them and hence confinement.  These have been seen in U(1),
SU(2) and SU(3) lattice gauge simulations thereby supporting this picture
of confinement.

In ordinary superconductivity, the primary issue is the spontaneous breaking
of the electromagnetic U(1) gauge symmetry (SSB)
signaled by the non-vanishing of
the vacuum expectation value of a charged field.
An immediate consequence
is that the curl of the vector potential is proportional to the curl
of the electric current known as  the London relation.
 The London relation is violated only near the boundaries of
superconducting material
within a distance of the order of the coherence length.
Combining the London relation with Maxwell's equations gives
mass to the electromagnetic field.  The Meissner effect and infinite
conductivity follow.

The theory of superconductivity entails (i) the identification of
what symmetry is broken and what are the relevant
coordinates, (ii) the mechanism that leads to the
the instability and hence  SSB, e.g. BCS theory, and then (iii) an
effective theory of the currents and fields in the broken phase, e.g.
Ginzburg-Landau [GL] theory,
which predicts the spatial consequences of the
broken symmetry\cite{t}.

Our goals are similar to these.
We hope to identify relevant coordinates and a corresponding
effective theory.   Experience has shown that an analytic approach
is very difficult.  The Villain form of the U(1) action
yields some analytic results\cite{bmk},
but the Wilson form for U(1) and SU(N) has
not.  Lattice simulations are
capable of building evidence for or against this picture and perhaps even be a
guide for analytic efforts.  The dual London relation found in
pure gauge U(1)\cite{shb} and SU(N)\cite{sbh,mes}
simulations gives evidence that an effective lattice dual Higgs theory
is a credible candidate mechanism for confinement.

Fields and currents exist only near the boundaries of dual superconductors
over a distance scale set by the London penetration depth $\lambda_d$.
It is these
spatially transient effects that provide spatial structures where
the local properties of dual superconductivity can be studied and that
is the focus of this work.  A dual
Abrikosov vortex is an example of such a structure.  The core is `normal
material' and the flux tube connecting quark and antiquark is a spacially
transient boundary effect.

\section{Abrikosov vortices in the lattice Higgs effective theory}

I would like to review briefly the salient features of the lattice
Higgs theory in the tree approximation.
It is relevant to note that in this model,
the London relation is a direct consequence of the
spontaneous breaking of the U(1) gauge theory.  We compare the simulations
of pure gauge theories with the dual of this model.

Consider the lattice action in a standard notation:
\begin{equation}
S = \beta \sum_{x,\mu > \nu} (1 - cos\;\theta_{\mu\nu}(x)) -
    \kappa \sum_{x,\mu}(\phi^{*}(x)e^{i\theta_{\mu}(x)}
                        \phi(x+\epsilon^{(\mu)}) + H.c.) +
    \sum_x V_{Higgs}(|\phi(x)|^2 ).
\label{e1}
\end{equation}

We are interested how the fields and currents respond to a magnetic
monopole anti-monopole pair. We obtain a classical solution by minimizing
the action.  Our initial configuration contains a single
closed magnetic monopole loop in the $x_3, x_4$ plane analogous to
the Wilson loop projector in the pure gauge simulations.
Our algorithm then rejects all updated links that change
the initial monopole configurations.
We use the method of simulated annealing,
slowly increasing $\beta$, holding  $ \lambda^2/a^2 = \beta/\kappa
= 1/e^2 \kappa$
constant, where $\lambda/a$ is the
London penetration
depth in lattice units and $a$ is the lattice spacing.

The electric current is given by
\begin{equation}
 \frac{a^3}{e \kappa} J^{e}_{\mu}(x) =  Im ( \phi^{*}(x)e^{i \theta_{\mu}(x)}
                                        \phi(x+\epsilon^{(\mu)})).
\label{e2}
\end{equation}
Writing the Higgs field $\phi(x) = \rho(x) e^{i \omega(x)}$,
the curl of the current is hence given by
\begin{eqnarray}
\lefteqn{ \frac{a^4}{e \kappa}(\Delta_{\mu} J^{e}_{\nu}(x)
                    - \Delta_{\nu} J^{e}_{\mu}(x))  =}\nonumber\\
             &&       \rho(x) \rho(x+\epsilon^{(\mu)})
                     \sin[ - \omega(x)+ \theta_{\mu}(x) +
                             \omega(x+\epsilon^{(\mu)})]+ \nonumber \\
               &&   \rho(x+\epsilon^{(\mu)})
                     \rho(x+\epsilon^{(\mu)}+\epsilon^{(\nu)})
                     \sin[ - \omega(x+\epsilon^{(\mu)})+
                              \theta_{\nu}(x+\epsilon^{(\mu)}) +
                     \omega(x+\epsilon^{(\mu)}+\epsilon^{(\nu)})]- \nonumber \\
               &&   \rho(x+\epsilon^{(\nu)})
                     \rho(x+\epsilon^{(\nu)}+\epsilon^{(\mu)})
                     \sin[ - \omega(x+\epsilon^{(\nu)})+
                              \theta_{\mu}(x+\epsilon^{(\nu)}) +
                    \omega(x+\epsilon^{(\mu)}+\epsilon^{(\nu)})]- \nonumber \\
               &&   \rho(x) \rho(x+\epsilon^{(\nu)})
                     \sin[ - \omega(x)+ \theta_{\nu}(x) +
                             \omega(x+\epsilon^{(\nu)})],
\label{e3}
\end{eqnarray}
where $\Delta_{\mu}$ is the lattice derivative.
Compare this with the electromagnetic field tensor
\begin{equation}
ea^2 F_{\mu \nu}  = \sin[  \theta_{\mu}(x)
                         + \theta_{\nu}(x+\epsilon^{(\mu)})
                         - \theta_{\mu}(x+\epsilon^{(\nu)})
                         - \theta_{\nu}(x)].
\label{e4}
\end{equation}
If the U(1) gauge symmetry is spontaneously broken, these two
quantities are equal which is the London relation.   To be more precise:
if (i) $\rho$ is nonvanishing and independent of position (absorb the
normalization into $\kappa$) and (ii)
\begin{equation}
 \sin[\theta + 2n\pi] \approx \theta,
\label{e4.1}
\end{equation}
 then
\begin{equation}
{\cal F}_{\mu \nu} \equiv  F_{\mu \nu}
   - \frac{a^2}{e^2 \kappa}(  \Delta_{\mu} J^{e}_{\nu}(x)
                            - \Delta_{\nu} J^{e}_{\mu}(x))
     = \frac{2\pi n}{e} \frac{1}{a^2}
    =  n e_m \frac{1}{a^2},
\label{e5}
\end{equation}
where ${\cal F}_{\mu \nu} a^2$ is called the fluxoid and $e_{m}$ is the
Dirac monopole charge.  Condition (i) is satisfied by the SSB potential
\begin{equation}
V_{Higgs}(|\phi(x)|^2) \Rightarrow |\phi(x)|^2 = 1,
\label{e5.1}
\end{equation}
and (ii) is satisfied if the lattice is a good approximation, yet still
allowing singular configurations when $n \neq 0$ giving the quantized
fluxoid.  (A Dirac string would act differently, adding  $2 \pi n$ to both
quantities.)
A `mexican hat potential' can also give the London relation
but it would be  violated if the plaquette is within the coherence length
of a superconducting-normal boundary where the Higgs field has a
non-vanishing gradient.

Eqn.\ref{e5}, together with Maxwell's equations, gives the Meissner
effect $ \vec{B} - \lambda^2 \nabla^2 \vec{B} = 0$, where
$ \lambda^2 = m_{\gamma}^{-2}$;
infinite conductivity
$\vec{E} = \lambda^2 \Delta_{4}\vec{J}$ (assuming
$\rho = 0$); and an Abrikosov vortex
\begin{equation}
B_z - \lambda^2(\vec{\Delta}\times \vec{J})_z =
 n \frac{e_m}{a^2}  \delta^2_{x_{\perp},\; 0}.
\label{e5.2}
\end{equation}

Fig.\ref{f1} shows the profile of the R.H.S.
 of eqn.(\ref{e3}) and eqn.(\ref{e4}) in the
directions perpendicular to the $5 \times 5$ magnetic monopole loop on a
$12^4$ lattice, with $\beta/\kappa = 1$. We used
the constrained form of the Higgs potential, eqn.(\ref{e5.1}),
which corresponds
to an extreme type II superconductor.
The graphs show the expected behavior: i.e. the equality of the two quantities
everywhere except at $r=0$ where they should differ by $2\pi$. There are
significant violations only at $r=0$ and $1$ due
 to the breakdown of Eqn. (\ref{e4.1}).
\begin{figure}
\vspace*{13pt}
\vspace*{1.6truein}		
\fcaption{Profile of $e a^2 F_{12}(x)$, eqn.(\ref{e4}), and $
-(a^4/e \kappa)(\Delta_{1} J^{e}_{2}(x)
                    - \Delta_{2} J^{e}_{1}(x))$, eqn.(\ref{e3}), as a
function of distance  perpendicular
to the plane of the monopole current loop in lattice units.}
\label{f1}
\end{figure}

Lattice artifacts can
violate the London and fluxoid quantization relations.   But ignoring these
controllable effects, the interior surface
spanned by the monopole loop is in the normal phase since the London relation
is violated there.
All other regions are superconducting.
This translates in the following
sections in which the Wilson loop projects out the normal phase  in the
plane spanned by the loop and the
remaining regions are a dual superconductor.

\section{Dual superconductivity in pure U(1) gauge theory}

We now turn to the dynamical simulations in pure U(1) gauge theory given by
the first term in the action, eqn.(\ref{e1}).   We have given evidence in
ref.\cite{shb} that a dual London relation is satisfied:
\begin{equation}
{\cal G}_{\mu \nu} \equiv
\frac{1}{2}\epsilon_{\mu \nu \sigma \tau} F_{\sigma \tau}
   - \lambda_{d}^2(  \Delta_{\mu} J^m_{\nu}(x)
                            - \Delta_{\nu} J^m_{\mu}(x))
     =  \frac{n e}{a^2} \delta_{x_{\mu},0} \; \delta_{x_{\nu},0}.
\label{e6}
\end{equation}
Magnetic monopoles are the current-carriers responsible for
the dual superconductivity.
These monopoles are defined in a 3-volume by the DeGrand-Toussaint\cite{dt}
construction, separating multiples of $2 \pi$ from the plaquette angle,
$(-\pi <  \bar{\theta}_{\mu \nu}(x) \leq \pi)$:
\begin{equation}
\theta_{\mu \nu}(x) = \bar{\theta}_{\mu \nu}(x) + 2\pi n_{\mu \nu}(x);\;\;\;
J^m_{\mu}(x)= \epsilon_{\mu \nu \sigma \tau}
(\bar{\theta}_{\sigma \tau}(x+\epsilon^{(\nu)}) -
 \bar{\theta}_{\sigma \tau}(x))
\label{e6.1}
\end{equation}
We associate $J^m_{\mu}(x)$ with  a link on the dual lattice,
making world lines which define a conserved current density.
The key idea is to measure the line integral of $J^m_{\mu}$ around a dual
plaquette, giving  $a^2(curl J^m)_{\mu \nu}$.   The correlation of this
with a Wilson loop, gives a signal for the solenoidal behavior
of the currents surrounding the electric flux between oppositely charged
particles.

\begin{figure}[htb]
\begin{picture}(200,120)(-100,0)
\newsavebox{\curla}
\savebox{\curla}{
\put(100,100){\line(1,0){40}}
\put(100,140){\line(1,0){40}}
\put(100,100){\line(0,1){40}}
\put(140,100){\line(0,1){40}}
\put(118,117){$\bullet$}
}
\newsavebox{\cord}
\savebox{\cord}{
\put(24,-8){z}
\put(-8,24){x}
\put(-20,-12){y}
\put(0,0){\line(1,0){24}}
\put(0,0){\line(0,0){24}}
\put(0,0){\line(-2,-1){12}}
}
\newsavebox{\curlb}
\savebox{\curlb}{
\put(100,140){\line(-2,-1){20}}
\put(100,140){\line(1,0){40}}
\put(140,140){\line(-2,-1){20}}
\put( 80,130){\line(1,0){40}}
\put(109,133){$\bullet$}
}
\newsavebox{\final}
\savebox{\final}{
\put(30,140){\usebox{\cord}}
\put(50,-2){\usebox{\curla}}
\put(50,42){\usebox{\curla}}
\put(48,-1){\usebox{\curlb}}
\put(72,11){\usebox{\curlb}}
\put(80,139){\rule{14.3mm}{0.6mm}}
\put(150,139){\rule{14.3mm}{0.5mm}}
\put(183,125){\line(0,1){42}}
\put(159,113){\line(0,1){42}}
\put(183,167){\line(-2,-1){24}}
\put(183,125){\line(-2,-1){24}}
\put(65,180){(a)}
\put(130,180){(b)} }
\put(-5,-80){\usebox{\final}}
\end{picture}

\fcaption{Operators for (a) $e a^2 F_{34}(x)$, and (b) $
(a^4/e_m)(\Delta_{1} J^{m}_{2}(x)
                    - \Delta_{2} J^{m}_{1}(x))$.}
\label{curl}
\end{figure}

Fig. \ref{curl} shows the lattice  operators for the electric
field and the curl of the magnetic monopole current.
The longitudinal electric
field, (a), is given by a $z,t$ plaquette which is
depicted by a bold line for fixed time.
The curl of the magnetic monopole current, (b), is built from four
3-volumes which appear as squares since the time dimension is not shown.
Passing through the center of each square is the link dual to the 3-volume.
One takes the monopole number $n$ in each 3-volume and associates the value
$n e_{m}$ with the corresponding dual link.
The `line integral' around the dual plaquette
completes the picture. Notice
from this construction that $\vec{E}$ and $\vec{\Delta}\times \vec{J}^m$ take
values at the same location
within the unit cell of the lattice, both are indicated by the
bold face line in the $z$ direction.  Both operators are in
defined on the same $z,t$ plaquette.

Fig.\ref{f3} shows the profile for the quantities making up the London
relation on the same plot as for the effective lattice Higgs case,
Fig.\ref{f1}.  We did 800 measurements on a $12^4$ lattice,
 $\beta = 0.95$, and a
$3\times3$ Wilson loop was used to project onto the $q,\bar{q}$ sector
using methods described in Ref.(\cite{shb}).  We did a $\chi^2$
fit of the London relation, eqn.(\ref{e6}),
 using the $r \neq 0$ points and determined
$\lambda_d = 0.49(3)$.  that determined the coefficient of the
delta function in eqn.(\ref{e6}), $n e = 1.07(7)$, compared to the expected
value $1/\sqrt{\beta} = 1.03$.  All these features correspond directly
to the classical solution of the lattice Higgs model reinterpreted as a
dual theory.
\begin{figure}
\vspace*{13pt}
\vspace*{1.6truein}		
\fcaption{Profile of $e a^2 F_{34}(x)$ and $
-(a^4/e_{m})(\Delta_{1} J^{m}_{2}(x)
                    - \Delta_{2} J^{m}_{1}(x))$ as a function of
distance  perpendicular
to the plane of the Wilson loop in lattice units.}
\label{f3}
\end{figure}

\section{Generalization to pure SU(2)and SU(3) gauge theories}

In ref.(\cite{sbh}) we applied these same techniques and showed that
dual Abrikosov vortices also occur in SU(2) pure gauge theory in the
maximal Abelian gauge.  Matsubara et.al.\cite{mes}
confirmed these results with
better statistics and
generalized them to SU(3).  The SU(2) link matrices are $U_{\mu}(x)$ and
the action is
\begin{equation}
 S = \beta \sum_{x,\mu > \nu}(1- \frac{1}{2}Tr U_{\mu \nu}(x)).
\label{e7}
\end{equation}
The maximal Abelian gauge is defined by maximizing the quantity
\begin{equation}
 R  =  \sum_{x,\mu}Tr[\sigma_3 U_{\mu}(x)\sigma_3 U_{\mu}^{\dagger}(x)].
\label{e8}
\end{equation}
The Abelian link angle is then taken as the phase of $[U_{\mu}(x)]_{11}$ and
the calculation can proceed with little change\cite{back2}.

Recently we have done the calculation at finite temperature in order to
check this picture on each side of the deconfining
phase transition\cite{ph} which we report here.
The results are shown in the confining phase, fig.\ref{f4},
$\beta = 2.28$  and  the deconfining phase, fig.\ref{f5},
$\beta = 2.40$ on a lattice $4 \times
17^2 \times 19$,  with 800 measurements for each case. Gauge fixing
required about 600 sweeps for each configuration.
\begin{figure}
\vspace*{13pt}
\vspace*{1.6truein}		
\fcaption{Same plot as in fig.\ref{f3} but for SU(2), in the confined phase}
\label{f4}
\end{figure}
\begin{figure}
\vspace*{13pt}
\vspace*{1.6truein}		
\fcaption{Same plot as in fig.\ref{f3},\ref{f4}
but for SU(2), in the deconfined phase.}
\label{f5}
\end{figure}
Fig.\ref{f4} shows an important difference from the U(1) case.  Whereas
in the U(1) case a linear combination of $E$ and $curl J^m$ for $r \neq 0$
can vanish giving the London relation, it is clearly not possible here. The
behavior of $-curl J^m$ does not match that of $E$.  We interpret this
discrepancy as a signal of a non-zero Ginzburg-Landau coherence
length, $\xi_d$.
 Whereas all evidence in the U(1) case points to the
extreme type II limit, i.e. the superconducting order parameter turning
on at the surface, the evidence here is that the order parameter turns on
with a behavior $tanh(x_{\perp}/\xi_d)$
The value of the coherence length is approximately
the radius where the London relation is restored.

The simulation shown in fig.\ref{f5} was identical to fig.\ref{f4}
except that $\beta$ was increased to $2.40$
putting us in the deconfining phase.
The dramatic decrease of $curl J^m$ results in the failure of
the dual superconductor interpretation as expected for the deconfining
phase.  Also $E$ falls more slowly with radius.

The behavior of all the SU(2) and SU(3) examples is similar to that shown in
fig.\ref{f4}.  The interesting conclusion is that
$\kappa_d \equiv \lambda_d/\xi_d \approx 1$ (no relation to Higgs $\kappa$)
For a  type II dual superconductor
 $\kappa_d > 1/\sqrt{2}$ and a type I otherwise.
These simulations indicate that the non-Abelian dual superconductors
lie at the borderline between type I and type II.  However a more systematic
study is needed to pin this down.

\section{External Electric Field}

Many interesting superconducting properties can be elucidated by studying
the properties of the material in the presence of a magnetic field. We
report a preliminary look at the corresponding problem of
dual superconductivity in the presence of an external electric field. The
Wilson loop provides such a field by projecting  a $q \bar{q}$
out of vacuum configurations.  But it may be interesting to see the
spontaneous breaking of translation invariance as the electric flux
forms a vortex rather than impose the vortex position at a given
location and with quantized flux.   Type I and type II dual
superconductors respond very differently to
background uniform external fields.

For a periodic U(1) lattice the sum of the plaquette angles over any
plane is identically zero.
A classical uniform electric field for U(1) on a particular time slice
 can still be obtained by
constraining one plaquette angle in each $z,t$ plane
to the value $(1/(N_zN_t)-1)\theta_c$.
Then the remaining $(N_z N_t-1)$
plaquette angles will take the common value
$\theta_c/(N_z N_t) = e a^2 F_{34}$ giving
a uniform electric field on all but one time slice. (N labels the lattice
dimensions.) Choosing $\theta_c = \pi$ gives the largest field.  For an
$8^4$ lattice and $\beta = 1$ for example, there would be enough electric
flux to form about 3 vortices in the $x,y$ plane in the dual superconducting
phase. This field configuration  can also be obtained by multiplying the
plaquette that was singled out in the action by a minus sign and such
configurations have been studied in non-Abelian
theories\cite{go}.  We avoided an
alternative method of imposing an external field by introducing a non-zero
equilibrium value locally for each plaquette angle since we eventually want
to see the field break translation invariance.
Turning on interactions for $\beta < 1$ brings up other interesting
features\cite{dh}.

Our goal here is to try to see a signal showing that $curl J^m$
responds to the external field.
The immediate problem is that the sum of $(curl J^m)_{x y}$
over the any $x,y$ plane is identically zero, and the sum of $E_z$ is not.
Yet
we expect the London relation to be satisfied.  The only possibility is that
translation invariance is broken which is expected since vortices segregate
the superconducting and normal phases.

We make the following rough ansatz that the local London relation is
due to a the alignment of local current loops in the external
field. This suggests that we truncate $curl J^m$ to include only
values $\pm 2,\pm 3,\pm 4$ representing the winding of the current around
the dual plaquette. We denote this $J^m_{\mu}(2+3+4)$.

Now we get a large signal for $curl J^m_{\mu}(2+3+4)$ as shown in
fig.\ref{f6} with a sign that agrees with the sign in fig.\ref{f3} for
the superconducting region, $r \neq 0$.  Further if we recalculate
fig.\ref{f3} with the truncated current, we find a large suppression at
$ r = 0$ and a moderate enhancement at the other points. In other words
this choice biases in favor of the dual superconducting phase.
\begin{figure}
\vspace*{13pt}
\vspace*{1.6truein}		
\fcaption{Average of $e a^2 F_{34}$ and $
-(a^4/e_{m})(\Delta_{1} J^{m}_{2}(2+3+4)
                    - \Delta_{2} J^{m}_{1}(2+3+4))$ on each time slice
for an external field $=\pi/64$ (the horizontal line).
The constrained plaquette is at $t=1$, the field is classical at
$t=2,8$ and $\beta=.99$ for $t=3-7$.}
\label{f6}
\end{figure}

\section{Acknowledgements}
I wish to thank D. Browne, L.-H. Chan, A. Di Giacomo,
Y. Peng, M. Polikarpov, H. Rothe,
G. Schierholz, V. Singh, J. Wosiek, and K. Yee for many discussions. We
are supported in part by the US Department of Energy grant
DE-FG05-91ER40617.

\end{document}